\documentclass[preprint,pteplogo]{ptephy_v2}%%%%%% to generate preprint number with ptep logo

\preprintnumber{XXXX-XXXX} %%% %%% Insert preprint number here
\usepackage{hyperref}

\usepackage{physics2}
\usephysicsmodule{ab} % automatic bracing
\usepackage[list-units=single, range-units=single]{siunitx} % SI units
\DeclareSIUnit{\dBm}{dBm} % definition of unit dBm
\usepackage{subcaption}

\begin{document}

\title{
    Feedback-Controlled Beam Pattern Measurement Method 
    Using a Power-Variable Calibration Source
    for Cosmic Microwave Background Telescopes
}

\author[1,2]{Haruaki Hirose}
\author[2,3,4]{Masaya Hasegawa}
\author[3]{Daisuke Kaneko}
\author[5]{Taketo Nagasaki}
\author[6,2]{Ryota Takaku}
\author[2,3,4]{Tijmen de Haan}
\author[7]{Satoru Takakura}
\author[3]{Takuro Fujino}

\affil[1]{
    Department of Physics, Graduate School of Engineering Science, 
    Yokohama National University,
    79-5 Tokiwadai, Hodogaya-ku, Yokohama, Kanagawa 240-8501, Japan
    \email{hirose-haruaki-bz@ynu.jp}}
\affil[2]{
    Institute of Particle and Nuclear Studies (IPNS), 
    High Energy Accelerator Research Organization (KEK),
    1-1 Oho, Tsukuba, Ibaraki 305-0801, Japan}
\affil[3]{
    International Center for Quantum-field Measurement Systems for Studies of the Universe and Particles (WPI-QUP), 
    High Energy Accelerator Research Organization (KEK),
    1-1 Oho, Tsukuba, Ibaraki 305-0801, Japan}
\affil[4]{
    The Graduate University for Advanced Studies (SOKENDAI),
    Shonan Village, Hayama, Kanagawa 240-0193, Japan}
\affil[5]{
    Accelerator Laboratory (ACCL), 
    High Energy Accelerator Research Organization (KEK),
    1-1 Oho, Tsukuba, Ibaraki 305-0801, Japan}
\affil[6]{
    Graduate School of Environmental, Life, Natural Science and Technology, 
    Okayama University, 
    3-1-1 Tsushima-Naka, Kita-ku, Okayama 700-8530, Japan}
\affil[7]{
    Department of Physics, Faculty of Science, 
    The University of Tokyo, 
    7-3-1 Hongo, Bunkyo-ku, Tokyo 113-0033, Japan}

\begin{abstract}%
We demonstrate a novel beam pattern measurement method 
for the side lobe characterization of cosmic microwave background telescopes.
The method employs a power-variable artificial microwave source under feedback control
from the detector under test on the telescope.
It enables us to extend the dynamic range of the beam pattern measurement 
without introducing nonlinearity effects from the detector. 
We conducted a laboratory-based proof-of-concept experiment, 
measuring the \textit{H}-plane beam pattern 
of a horn antenna coupled to a diode detector at \qty{81}{\GHz}.
We gained an additional dynamic range of \qty{60.3}{\decibel} attributed to the feedback control.
In addition, we verified the measurement by comparing it
with other reference measurements obtained using conventional methods.
The method is also applicable to general optical measurements
requiring a high dynamic range
to detect subtle nonidealities in the characteristics of optical devices.
\end{abstract}

\subjectindex{F10, F14, H52}

\maketitle

\section{Introduction}
\label{sec:introduction}

Precise measurements of degree-scale B-mode polarization patterns 
in the cosmic microwave background (CMB) provide
information about inflation in the early universe \cite{Seljak1997, Kamionkowski1997}.
The inflationary B-mode signal has not been detected yet,
and further observations are ongoing or planned 
with ground-based or space-borne telescopes
(e.g., \cite{SPT3G2018, SO2019, BK2022, CMBS42022, LiteBIRD2023}).
These experiments aim to achieve sufficient statistical sensitivity to detect the B-mode signal 
by using arrays of highly sensitive detectors such as transition edge sensor (TES) bolometers,
combined with multi-year observation campaigns.
However, controlling systematic errors is crucial.
A major source of systematic errors is the leakage of galactic foregrounds through side lobes,
undesired gain outside the telescope’s line of sight.
These foregrounds are bright enough to leak into the B-mode signal,
leading to a significant systematic error.

To avoid such contamination, it is necessary to precisely characterize the side lobes,
in addition to using a telescope with inherently low side lobe levels.
Comprehensive characterization is required
to verify that the side lobe level meets design specifications
and to enable accurate foreground removal during data analysis.
Even extremely faint side lobes 
(e.g., those at \qty{30}{\decibel} or more below the main lobe) 
must be identified prior to science observation;
otherwise, they may introduce systematic errors in the B-mode search
over long-term observations. 
In particular, space-borne telescopes must rigorously characterize their side lobes
during pre-launch laboratory testing.
For example, \textit{LiteBIRD}, a future CMB observation satellite, 
requires far side lobe calibration down to \qty{-56}{\decibel} \cite{LiteBIRD2023}.
Thorough side lobe characterization during pre-deployment or pre-launch laboratory testing is important for future CMB telescopes 
to ensure high-quality B-mode observations.

Accurate characterization of such faint side lobes demands
a high dynamic range measurement of the beam pattern, 
the telescope's angular gain profile.
However, the TES bolometers employed in current- and next-generation CMB telescopes
have dynamic ranges that typically fall short of this requirement.
The dynamic range of a detector is defined as
\begin{equation}
    \mathrm{DR}_\mathrm{dB}[P_\mathrm{r}] 
        = 10 \log_{10} \frac{\max P_\mathrm{r}}{\Delta P_\mathrm{r}} \, \unit{\decibel},
    \label{eq:dynamic_range}
\end{equation}
where $P_\mathrm{r}$ denotes the power received by the detector 
and $\max P_\mathrm{r}$ and $\Delta P_\mathrm{r}$ represent 
the maximum measurable power%
\footnote{Commonly defined as the \qty{1}{\decibel} compression point.}
and the minimum detectable power
within its linear response range, respectively.
The maximum measurable power is significantly lower than the saturation power $P_\mathrm{sat}$ 
(typically on the order of \unit{\pico \watt} in TES bolometers used in CMB telescopes
\cite{Stevens2020, Jaehnig2020, BK2022, deHaan2024}) 
due to detector nonlinearity, 
and the minimum detectable power is determined by the noise-equivalent power, $\mathrm{NEP}$
(typically tens of \unit{\atto \watt \sqrt{\second}} level \cite{Dutcher2024}),
and signal integration time $t$.
For example, assuming $\max P_\mathrm{r} < 0.1 P_\mathrm{sat}$ 
and $\Delta P_\mathrm{r} \sim \mathrm{NEP} / \sqrt{t}$ 
and using the typical values above with an integration time of $t = \qty{1}{\second}$,
the dynamic range is estimated to be less than \qty{40}{\decibel}.
It is also difficult to extend the dynamic range by using signals in the nonlinear region in practice.
Correcting distorted signals requires characterizing and understanding the actual nonlinearity of each TES bolometer on the telescope \cite{deHaan2024}.
The limited dynamic range hinders the detection of faint side lobes
with TES bolometers installed in the telescopes.

CMB telescopes commonly measure their beam patterns by scanning bright, point-like static sources in the sky
(e.g., planets \cite{WMAP2003a, PB2014, Planck2016}).
Repeating and averaging multiple measurements over the observation period 
improves the signal-to-noise ratio and extends the dynamic range.
However, achieving a high dynamic range such as \qty{60}{\decibel} using this method is impractical, 
as a \qty{10}{\decibel} increase in dynamic range requires statistically \num{100} times more data.
Additionally, long-term instabilities in the source or instruments 
can introduce significant systematic errors.
A more practical approach is to combine multiple measurements using sources of varying intensities.
Examples include celestial bodies such as planets, the Sun, and the Moon \cite{WMAP2003b, Planck2016, QUIET2013, CLASS2024}.
The BICEP/\textit{Keck} collaboration employed an artificial microwave source 
with several selectable intensity levels using built-in attenuators,
achieving a dynamic range of \qty{\sim 70}{\decibel}
in their on-site side lobe calibration \cite{BK2015, BK2022}.%
\footnote{
    Note that their TES bolometers had a special operation mode for calibration only, 
    in which the bolometers were biased at a higher transition point 
    and extended the saturation power and dynamic range 
    \cite{BK2015, BK2022}.}
In this multi-source approach, however, the TES bolometer's nonlinearity can distort the united beam pattern 
when stitching multiple measurements together.
This concern is particularly significant in laboratory testing for pre-deployment or pre-launch characterization.
Since bolometer conditions, such as optical loading or noise level, 
in the laboratory may differ from those in situ,
the bolometers may exhibit greater nonlinearity than expected.

\begin{figure}[!htbp]
    \centering
    \includegraphics[width=\columnwidth]{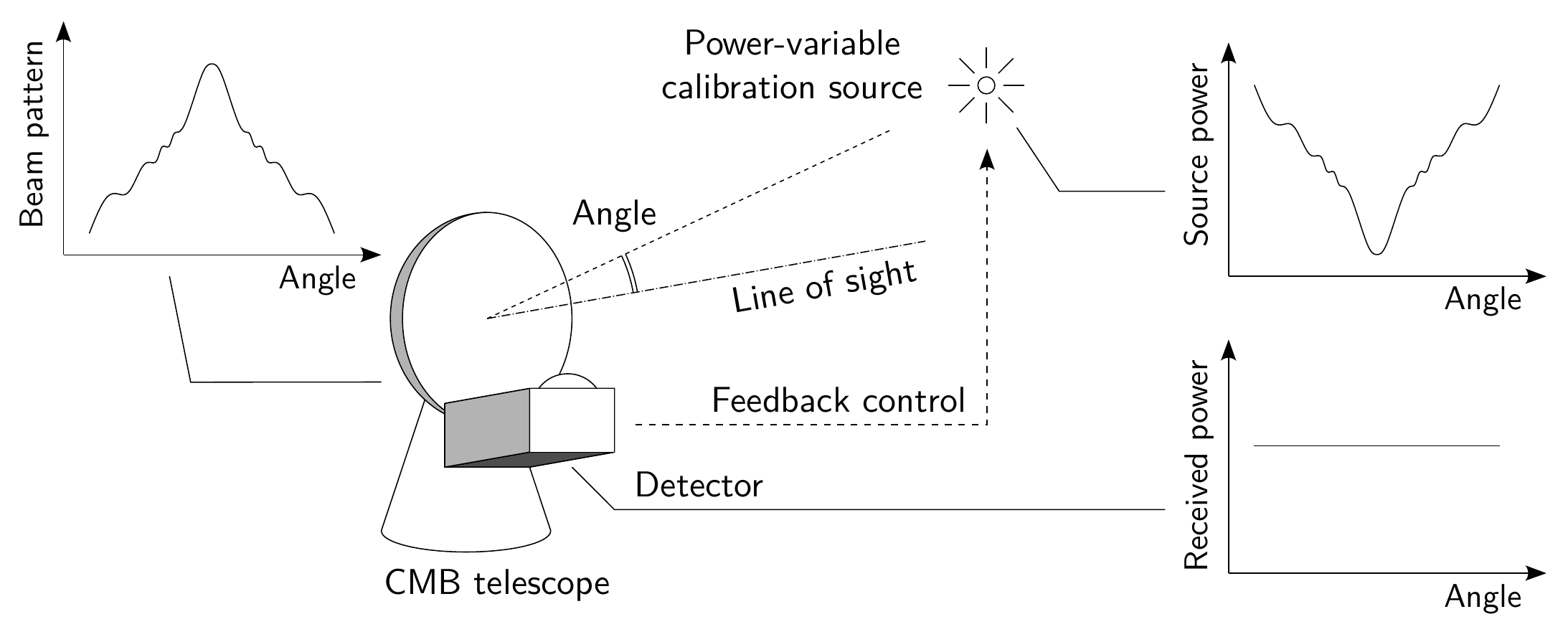}
    \caption{
        Concept of the feedback-controlled beam pattern measurement method.
        The beam pattern measurement is conducted
        using the power-variable calibration source under feedback control from 
        the detector under test on the telescope.
        The feedback control keeps the received power by the detector constant 
        to avoid detector nonlinearity during the measurement.
    }
    \label{fig:concept} 
\end{figure}

This study presents an advanced measurement method for high dynamic range beam pattern characterization.
The concept of the proposed method is illustrated in Fig. \ref{fig:concept}.
The method employs a power-variable artificial microwave source.
Feedback control from the detector under test on the telescope is applied to the source, 
dynamically adjusting the source power during measurement
to maintain the received power at a constant level
within the detector's linear response range.
This approach renders the measurement results insensitive to detector nonlinearity.
It effectively extends the dynamic range of the beam pattern measurement 
regardless of the detector's intrinsic dynamic range.

Section \ref{sec:methodology} describes the methodology of the proposed approach.
Section \ref{sec:poc} presents a laboratory-based proof of concept, 
demonstrating that the method achieves high dynamic range beam pattern measurement as intended.
Finally, Sec. \ref{sec:conclusion} summarizes the conclusions and discusses the prospects of the method. 
We also mention broader applications of the method to general optical measurements
beyond beam pattern characterization of CMB telescopes.

\section{Methodology}
\label{sec:methodology}

\begin{figure}[!htbp]
    \centering
    \includegraphics[width=\columnwidth]{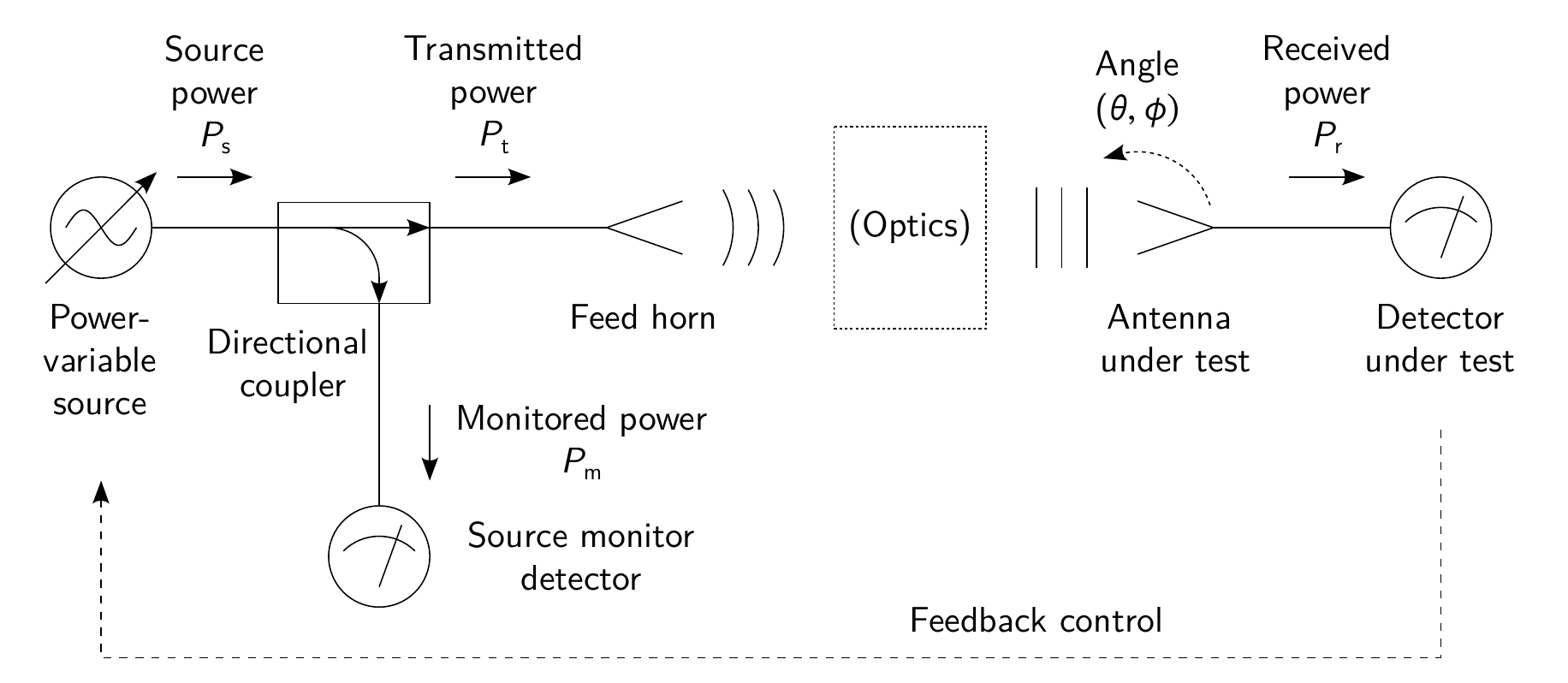}
    \caption{
        General model of the feedback-controlled beam pattern measurement system.
        The microwave generated by the power-variable source
        propagates from the feed horn to free space, 
        passes the optics (e.g., a compact range or a far-field distance),
        and illuminates the antenna under test coupled to the detector under test.
        The source monitor tracks the source power through the directional coupler.
        The closed-loop feedback control from the detector under test
        adjusts the source power during the measurement.
    }
    \label{fig:model} 
\end{figure}

Figure \ref{fig:model} shows a high-level illustration of the general system 
used for a beam pattern measurement in the proposed method.
A microwave is generated by a power-variable source
and is transmitted from a feed horn into free space.
After passing through an optical system (e.g., a compact range) or over a far-field distance,
the microwave is well-approximated by a plane wave and illuminates the antenna under test (AUT) 
at an angle $(\theta, \phi)$.
The AUT is coupled with an amplitude detector
(referred to as the detector under test [DUT])
to measure the received power $P_\mathrm{r}$.
A second detector, referred to as the ``source monitor,'' is connected to the source via a directional coupler.
The source monitor tracks the source power $P_\mathrm{s}$ during the measurement
by sampling it through the coupler and measuring the monitored power $P_\mathrm{m}$.
This is essential
because the source may exhibit nonlinearity or lack calibration in practice.

The system incorporates closed-loop feedback control from the DUT to the source.
The control loop dynamically adjusts the source power during measurement
to maintain the received power at a constant target level, $P_\mathrm{r,target}$, 
set within the DUT's linear response range.
This measurement scheme excludes distortion due to DUT nonlinearity from the measurement
by keeping the received power constant during the measurement.
In effect, the method transfers the linearity requirement
from the DUT to the source monitor detector. 
By selecting a source monitor detector with better linearity than the DUT, 
the overall linearity of the measurement is improved.
This approach is particularly helpful when the DUT is a nonlinear detector, such as a TES bolometer.

The adjustable power range of the source is constrained 
by the linear response range of the source monitor 
or the emissible power output range of the source.
When the source power reaches the upper limit of the adjustable power range, 
the measurement is conducted within the DUT's own dynamic range, 
as in a conventional static source measurement.

This method requires appropriately determining
the minimum source power, $\min P_\mathrm{s}$,
and the coupling factor, $C$, of the directional coupler.
Assuming the loss of the coupler is negligible, the requirements are
\begin{subequations}
    \begin{gather}
        \min P_\mathrm{s} \leq \frac{1}{\eta} P_\mathrm{r,target}, \\
        C \leq -10 \log_{10} \ab( \eta \, \frac{\Delta P_\mathrm{m}}{P_\mathrm{r,target}} ) 
            \, \unit{\decibel} \label{eq:coupling},
    \end{gather}
\end{subequations}
where $\eta = {P_\mathrm{r}(\theta_0, \phi_0)}/{P_\mathrm{t}(\theta_0, \phi_0)}$ 
represents the total optical eﬃciency from the feed horn to the DUT at the beam center,
where $P_\mathrm{t}$ is the transmitted power from the feed horn
and $\Delta P_\mathrm{m}$ is the minimum detectable power of the source monitor.
The optical eﬃciency depends on the optical system configuration.
These requirements ensure 
that the source and monitored powers remain sufficiently high at the beam center.
The coupling factor is optimal when equality holds in Eq. \ref{eq:coupling},
where the total dynamic range of the beam pattern measurement is maximized.

The normalized power beam pattern at an angle $(\theta, \phi)$ is given by
\begin{equation}
    G_\mathrm{n}(\theta, \phi) = 
        \frac{1}{G_0}
    \frac{P_\mathrm{r}(\theta, \phi)}{P_\mathrm{m}(\theta, \phi)},
    \label{eq:beam}
\end{equation}
where $G_0 = P_\mathrm{r}(\theta_0, \phi_0) / P_\mathrm{m}(\theta_0, \phi_0)$ is the normalization factor
and $(\theta_0, \phi_0)$ denotes the angle at which the beam pattern reaches its maximum 
(typically at the beam center).
By the feedback control, the beam pattern measurement obtains an additional dynamic range,
in which the systematic error due to the DUT's nonlinearity is excluded.
This additional dynamic range is gained independently of the DUT's own dynamic range.
Assuming that the coupling factor is optimized as previously described
and that the adjustable power range of the source is limited by the linearity of the source monitor,
the additional dynamic range corresponds to that of the source monitor detector, $\mathrm{DR}_\mathrm{dB}[P_\mathrm{m}]$, 
which is defined consistently with Eq. \ref{eq:dynamic_range}.
Therefore, a detector with good linearity and a high dynamic range is suitable 
as a source monitor in high dynamic range measurements.
The whole measurement can be performed independently of the DUT's dynamic range,
provided that the source monitor offers a higher dynamic range than required.

In some cases, the source may saturate in a high-power region and run short of power during the feedback control.
When this happens, the adjustable power range of the source is limited by its maximum emissible power, $\max P_\mathrm{s}$, 
rather than by the maximum measurable power of the source monitor.
Then, the additional dynamic range gained by the feedback control reduces to
$10 \log_{10} \ab( \eta \frac{\max P_\mathrm{s}}{P_\mathrm{r,target}}) \, \unit{\decibel}$.
However, commercial artificial microwave sources can typically emit \qty{10}{\dBm} or more, 
which is sufficient in most cases, including the demonstration shown in Sec. \ref{sec:poc}.

\section{Proof-of-concept test}
\label{sec:poc}

To demonstrate proof of concept, we applied this method to a laboratory optical system 
and measured the \textit{H}-plane beam pattern%
\footnote{
    The one-dimensional beam pattern 
    taken in the plane that contains the main lobe axis and the magnetic field vector.
    We chose the \textit{H}-plane over the \textit{E}-plane 
    because the horn antenna's side lobe level is lower in the \textit{H}-plane, 
    which is suitable for demonstrating a high dynamic range measurement.
    }
of a W-band (\qtyrange{75}{110}{\GHz}) standard gain pyramidal horn antenna (Millitech SGH-10-RP000)
coupled to a commercial diode amplitude detector (Eravant SFD-753114-103-10SF-P1).
We set the target dynamic range of the measurement at \qty{60}{\decibel},
which satisfies \textit{LiteBIRD}'s calibration requirement of \qty{56}{\decibel} \cite{LiteBIRD2023}.

\subsection{Test setup}

\begin{figure}[!htbp]
    \centering
    \includegraphics[width=0.90\columnwidth]{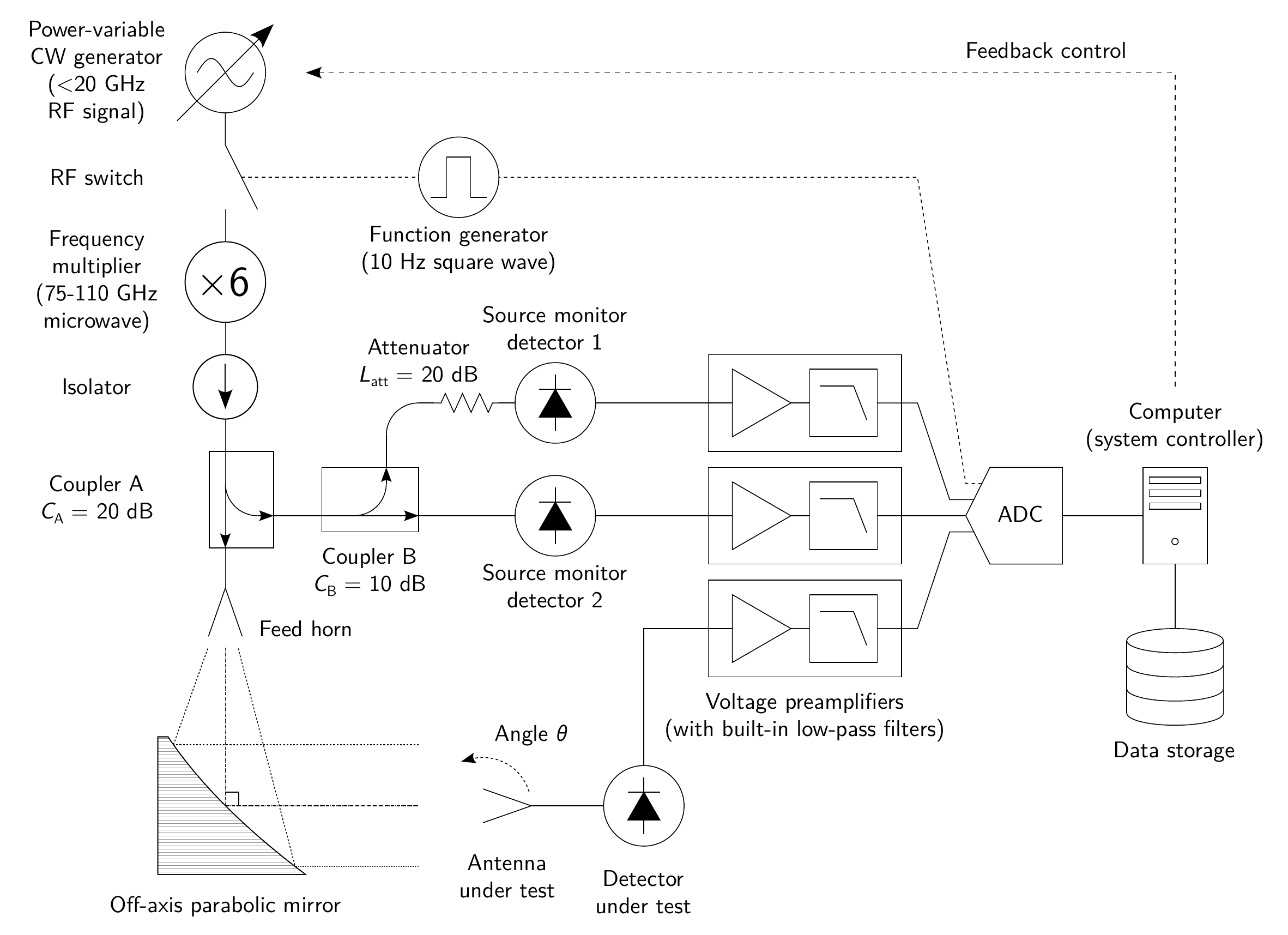}
    \caption{
        Block diagram of the feedback-controlled beam pattern measurement system 
        used for the proof-of-concept demonstration. 
        The power-variable continuous wave (CW) generator and frequency multiplier 
        produce a W-band (\qtyrange{75}{110}{\GHz}) microwave,
        which is pulse-modulated at \qty{10}{\Hz} by the RF switch. 
        The \qty{20}{\decibel} directional coupler (Coupler A)
        transmits the microwave to the feed horn attached to the through port,
        and two source monitor detectors (diode amplitude detectors) 
        via the \qty{10}{\decibel} directional coupler (Coupler B) 
        attached to the coupling port.
        One of the source monitor detectors is given a \qty{30}{\decibel} attenuation
        by Coupler B and the \qty{20}{\decibel} attenuator.
        The feed horn feeds the compact range optics, which consists of the off-axis parabolic mirror,
        and illuminates the antenna under test connected to the detector under test.
        All detector outputs are amplified, low-pass-filtered, digitized, 
        and processed in the system-control computer, which also runs the feedback control.
    }
    \label{fig:system} 
\end{figure}

\begin{figure}[!htbp]
    \centering
    \begin{subfigure}{0.45\columnwidth}
        \includegraphics[width=\columnwidth]{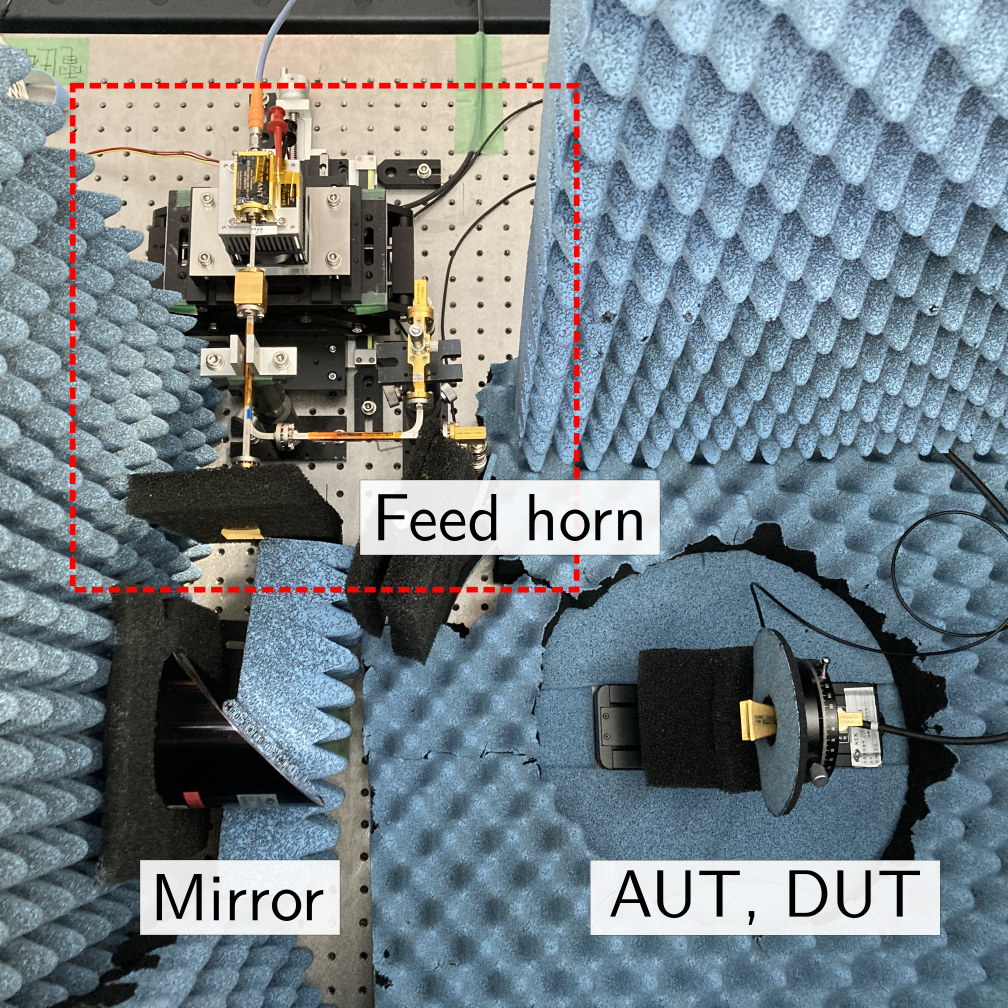}
        \label{fig:system_photo_left}
    \end{subfigure}
    \hfill
    \begin{subfigure}{0.45\columnwidth}
        \includegraphics[width=\columnwidth]{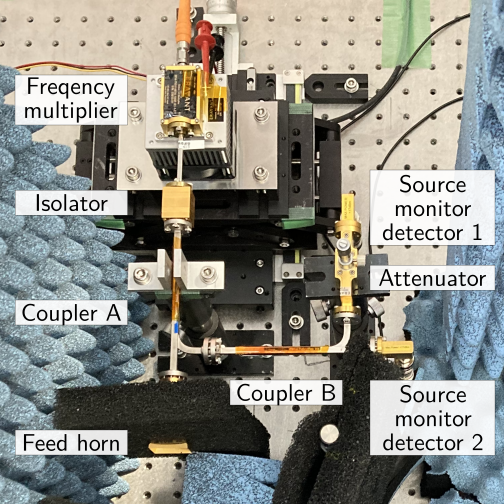}
        \label{fig:system_photo_right}
    \end{subfigure}
    \caption{
        Photographs of the feedback-controlled beam pattern measurement system.
        Left: Compact range.
            It was fully shielded by microwave absorbers during the measurement
            to suppress unexpected stray lights.
            Some absorbers are temporarily removed for the photograph.
            The antenna under test (AUT) and detector under test (DUT) 
            are mounted on an automatic rotation stage
            to change the AUT angle.
        Right: Close-up of the source and source monitor
            (the part framed with red dashed lines in the left panel).
            The continuous wave generator and RF switch are not shown in this photograph.
    }
    \label{fig:system_photo}
\end{figure}

We constructed a feedback-controlled beam pattern measurement system in our laboratory,
as shown in Figs. \ref{fig:system} and \ref{fig:system_photo}.
A power-variable continuous wave (CW) generator (Agilent Technologies E8247C) 
combined with a W-band frequency multiplier (Eravant SFA-753114616-10SF-E1-1-ET) was used as the source.
The generated microwave was pulse-modulated at \qty{10.0}{\Hz}
using an RF switch (Analog Devices ADRF5045-EVALZ), 
driven by a square wave from a function generator.
An isolator was attached to the output port of the multiplier
for protection from unexpected microwave reflection.
The source monitor was connected to the source
via a \qty{20}{\decibel} directional coupler
(Coupler A in Fig. \ref{fig:system}).
The coupling factor satisfies Eq. \ref{eq:coupling}.
The source monitor comprised two diode amplitude detectors (Eravant SFD-753114-103-10SF-P1, the same model as the DUT), 
which are common linear microwave detectors 
with output voltage proportional to the incident power.
Their typical sensitivity and maximum measurable power in their linear range were \qty{1000}{\mV/\mW} and \qty{-30}{\dBm}, respectively.%
\footnote{
    Based on the specifications and calibration provided by the manufacturer.
    While the typical input power is specified at \qty{-20}{\dBm},
    we conservatively assumed the maximum measurable power to be \qty{-30}{\dBm}
    to ensure sufficient linearity.
    }
The noise level was \qty{\sim 200}{\nV \sqrt{\s}} when combined with the voltage preamplifier described later,
allowing the detection of a power of approximately \qty{-70}{\dBm} at a minimum
over a signal integration time of \qty{10}{\second}.
Because a single diode detector cannot monitor
source power variation over a \qty{60}{\decibel} dynamic range,
two detectors were combined:
one (Source monitor detector 1) included a \qty{30}{\decibel} attenuation via another directional coupler (Coupler B) and an attenuator,
while the other (Source monitor detector 2) was unattenuated.
The feed horn (a W-band standard gain pyramidal horn antenna, Millitech SGH-10-RP000, the same model as the AUT)
transmitted the microwave across a compact range.
The compact range consisted of an off-axis parabolic mirror (Edmund Optics \#36602),
with a diameter of \qty{101.6}{\mm}, offset angle at \qty{90}{\degree}, 
and an effective focal length of \qty{152.4}{\mm}.
The AUT was coupled to the DUT and mounted within the compact range
on an automatic rotation stage that changed the angle of the AUT azimuthally.
The DUT's characteristics are identical to those of the source monitor detectors described above.
The distance from the mirror to the AUT's aperture was about \qty{390}{\mm}.
The optical efficiency of the compact range was $\eta \sim \qty{-10}{\decibel}$.
We carefully aligned the entire optics using visible lasers. 
The \textit{H}-plane of the rectangular aperture of the feed horn and the AUT are aligned horizontally.
The aperture center of the AUT was also aligned with the rotation axis of the rotation stage
so that the aperture is illuminated on the optical axis of the compact range at any angle.
The alignment accuracy of the position and the horizontal angle of each optical component 
is estimated to be \qty{\sim 1}{\mm} and \ang{\sim 2}, respectively.

Signals measured by the two source monitor detectors and the DUT 
were amplified by voltage preamplifiers,
filtered by built-in \qty{100}{\Hz} low-pass filters within the preamplifiers, 
and simultaneously read out by a multichannel analog-to-digital converter (ADC)
at a sampling frequency of \qty{1}{\kHz}.
The voltage gain of the preamplifiers was set to \num{2e3}.
To obtain the voltage amplitude of the modulated signal measured by each detector,
coherent demodulation was performed using custom software on the system-control computer,
employing the reference signal from the function generator to the ADC.
The demodulated signals from the two monitor detectors were merged 
by scaling with the ``intermonitor coefficient,''
defined as the ratio of their sensitivities to the source power. 
Signals from Source monitor detector 2 were used when its received power was within its linear response range
(less than \qty{-30}{\dBm});
otherwise, signals from Source monitor detector 1 were used.
The intermonitor coefficient was determined
by feeding a test microwave from the source to the two detectors
and measuring the ratio of their output voltages.
The determined intermonitor coefficient was $c_\mathrm{m} = \num{8.484e-4} = \qty{-30.7}{\decibel}$ at \qty{81}{\GHz}.

\subsection{Procedure}

\begin{figure}[!htbp]
    \centering
    \includegraphics[width=0.8\columnwidth]{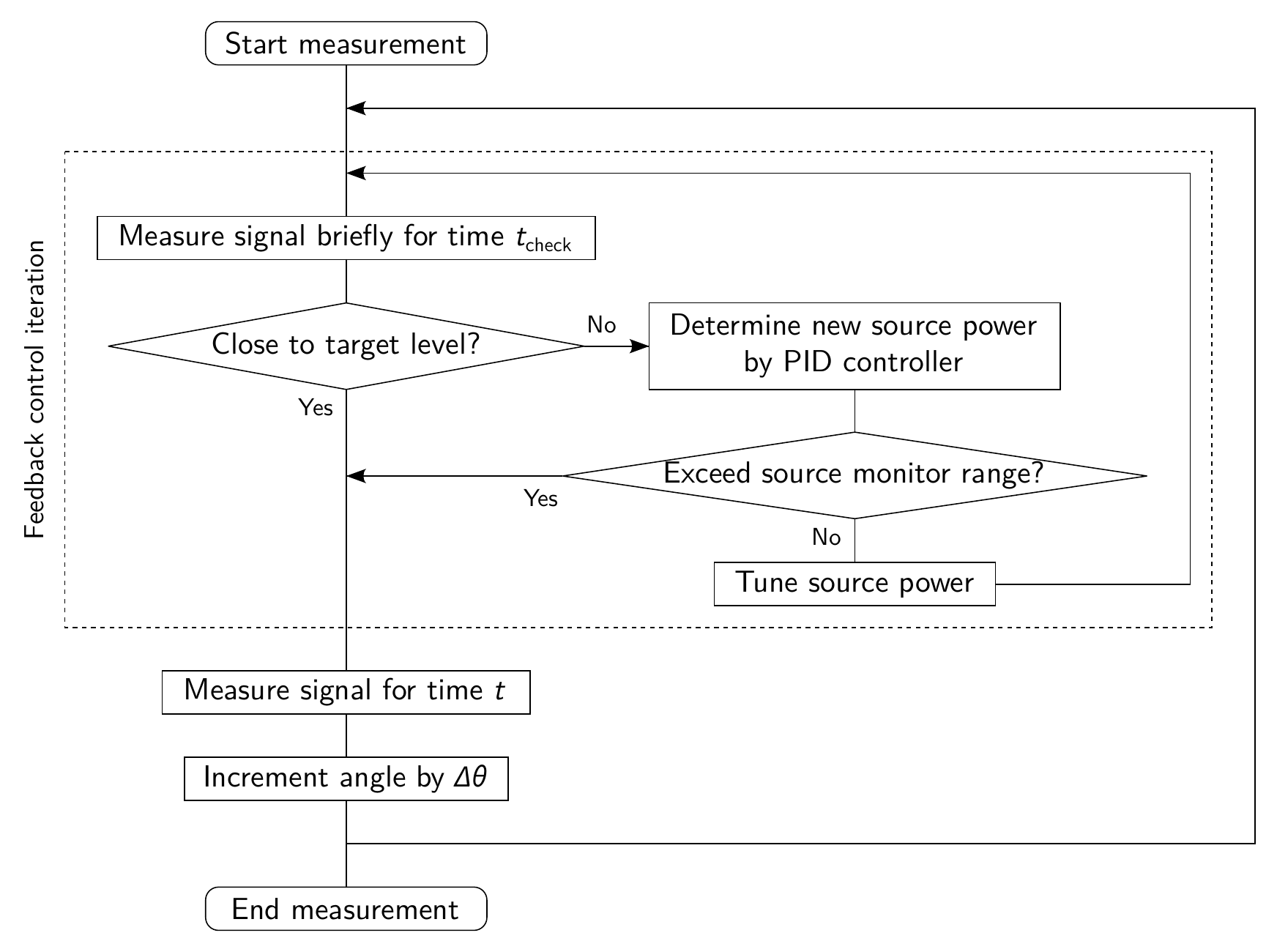}
    \caption{
        Flowchart of the routine 
        in the feedback-controlled beam pattern measurement.
        The routine was repeated until all the angular points were covered.
        The subroutine boxed in the dashed lines
        is the closed-loop feedback iteration adjusting the source power.
    }
    \label{fig:flowchart} 
\end{figure}

The beam pattern was measured at discrete angular points 
over the angular range of \qtyrange[range-units=repeat]{-10}{180}{\degree} to the beam center.
The angular interval (i.e., the angular resolution of the measured beam pattern) 
was $\Delta \theta = \ang{1.0}$.
At each point, the feedback control tuned the source power 
according to the routine shown in Fig. \ref{fig:flowchart}.
The DUT signal was briefly measured ($t_\mathrm{check} = \qty{1}{\second}$) to assess its current amplitude.
The source power was adjusted by a PID controller 
until the DUT signal amplitude reached
within \qty{\pm 10}{\percent} of the target signal level
that corresponds to the target level of the received power at the DUT.
We set the target level at $P_\mathrm{r, target} = \qty{-50}{\dBm}$,
intentionally set much lower than the DUT's maximum measurable power of \qty{-30}{\dBm} 
to demonstrate a test with a low dynamic range DUT.
The PID controller gains were manually tuned in advance.
When the monitored power reaches the maximum measurable power of the source monitor
(\qty{-30}{\dBm} at Source monitor detector 1),
the system terminates feedback control and maintains the source power at this limit.
This tuning algorithm was performed using custom software on the system-control computer.
After adjusting the source power, signals from the source monitor and the DUT were measured 
over a signal integration time of $t = \qty{10}{\second}$.

The beam pattern was estimated using Eq. \ref{eq:beam}.%
\footnote{
    The calculation used measured detector voltages, 
    whereas Eq. \ref{eq:beam} is expressed in terms of the measured powers. 
    The two formulations are equivalent.}
The main lobe in the angular range of $|\theta| \leq \ang{10}$ 
was fitted using a model described in Ref. \cite{Stutzman1998}
to determine the normalization factor and to correct the angle of the beam center.%
\footnote{
    We fitted only the main lobe region, in which the measurement and the model agreed sufficiently.
    The model cannot precisely describe the actual side lobes,
    mainly due to diffraction effects.
    }%
\footnote{
    The angle correction of the beam center was performed 
    for possible misalignment of the AUT with respect to the optical axis of the compact range. 
    }
The uncertainty at each point of the beam pattern was derived
by propagating the uncertainties
in the DUT voltage amplitude, source monitor voltage amplitude, 
normalization factor, and intermonitor coefficient.

\subsection{Results and discussion}

\begin{figure}[!htbp]
    \centering
    \includegraphics[width=0.90\columnwidth]{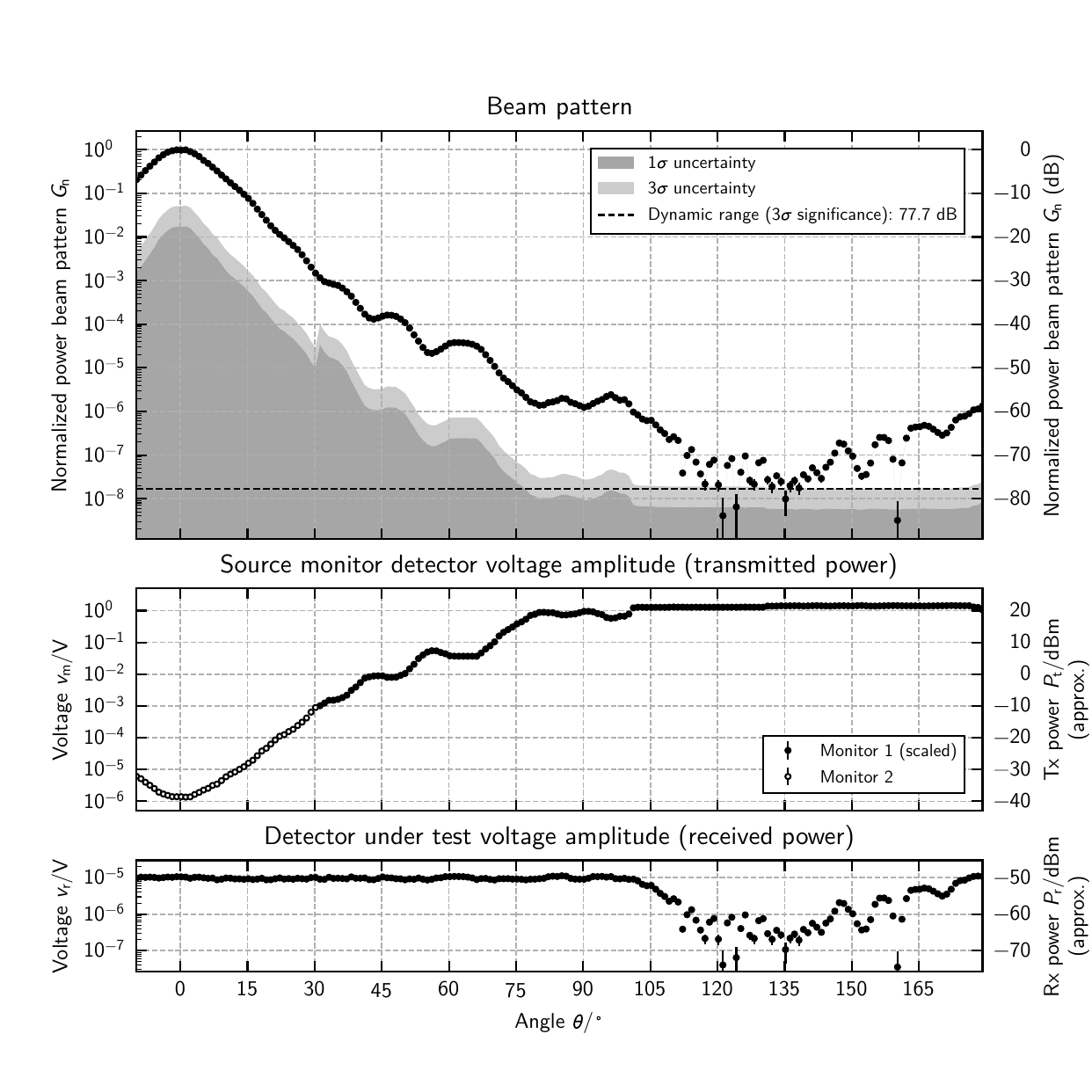}
    \caption{
        \textit{H}-plane beam pattern of the W-band standard gain pyramidal horn antenna 
        at \qty{81}{\GHz}
        measured in the feedback-controlled measurement method (top), 
        with the corresponding voltage amplitudes of the source monitor detectors (middle) 
        and the detector under test (bottom).
        The estimated transmitted (Tx) power from the feed horn 
        and the received (Rx) power at the detector under test are also shown 
        in the right axis of the middle and bottom panels, respectively.
        The dark- and light-shaded areas in the top panel show
        the magnitude of the $1\sigma$ and $3\sigma$ uncertainty. 
        The dashed line in the top panel shows 
        the minimum detectable level at the $3\sigma$ significance,
        indicating the achieved total dynamic range of the measurement.
        The filled and unfilled points in the middle panel are from 
        Source monitor detectors 1 and 2, respectively. 
        The former is scaled using the intermonitor coefficient 
        (the ratio of their sensitivities to the source power)
        and is stitched to the latter.
    }
    \label{fig:beam_feedback}
\end{figure}

Figure \ref{fig:beam_feedback} shows the \textit{H}-plane beam pattern at \qty{81}{\GHz}
measured using the feedback-controlled method.
The corresponding source power, measured by the source monitor,
and the received power at the DUT are also shown.
During the measurement, the feedback control adjusted the source power over a range of \qty{60.3}{\decibel},
as measured by the source monitor.
As a result, the system successfully maintained the received power at the DUT 
near the target level of \qty{-50}{\dBm} throughout the measurement.
The measurement gained an additional dynamic range of \qty{60.3}{\decibel} by using the feedback-controlled method,
achieving a total dynamic range of \qty{77.7}{\decibel} by including the DUT's dynamic range.
We performed the same measurement at several W-band frequencies
and achieved similar dynamic ranges.
These high dynamic ranges enabled the detection of very faint side lobes.

To highlight the advantages of the proposed method, 
the same beam pattern was measured without feedback control, 
that is, with the source power fixed at certain levels.
Two cases were demonstrated:
one using a static source with appropriate intensity that did not saturate the DUT
and another with higher intensity that intentionally saturated the DUT.
The received power at the DUT at the beam center was 
approximately \qtylist{-50;0}{\dBm} in each case, respectively.%
\footnote{\qty{0}{\dBm} stays safely below the \qty{17}{\dBm} handling power limit of the DUT.}
These values correspond to transmitted powers from the feed horn of 
\qtylist{-40;10}{\dBm}, respectively.
In the high-intensity source measurement, the voltage gain of the DUT's preamplifier was reduced
to avoid exceeding the maximum voltage range of both the preamplifier and the ADC;
otherwise, the exceeding signals would be clipped at their voltage limit.

\begin{figure}[!htbp]
    \centering
    \includegraphics[width=0.90\columnwidth]{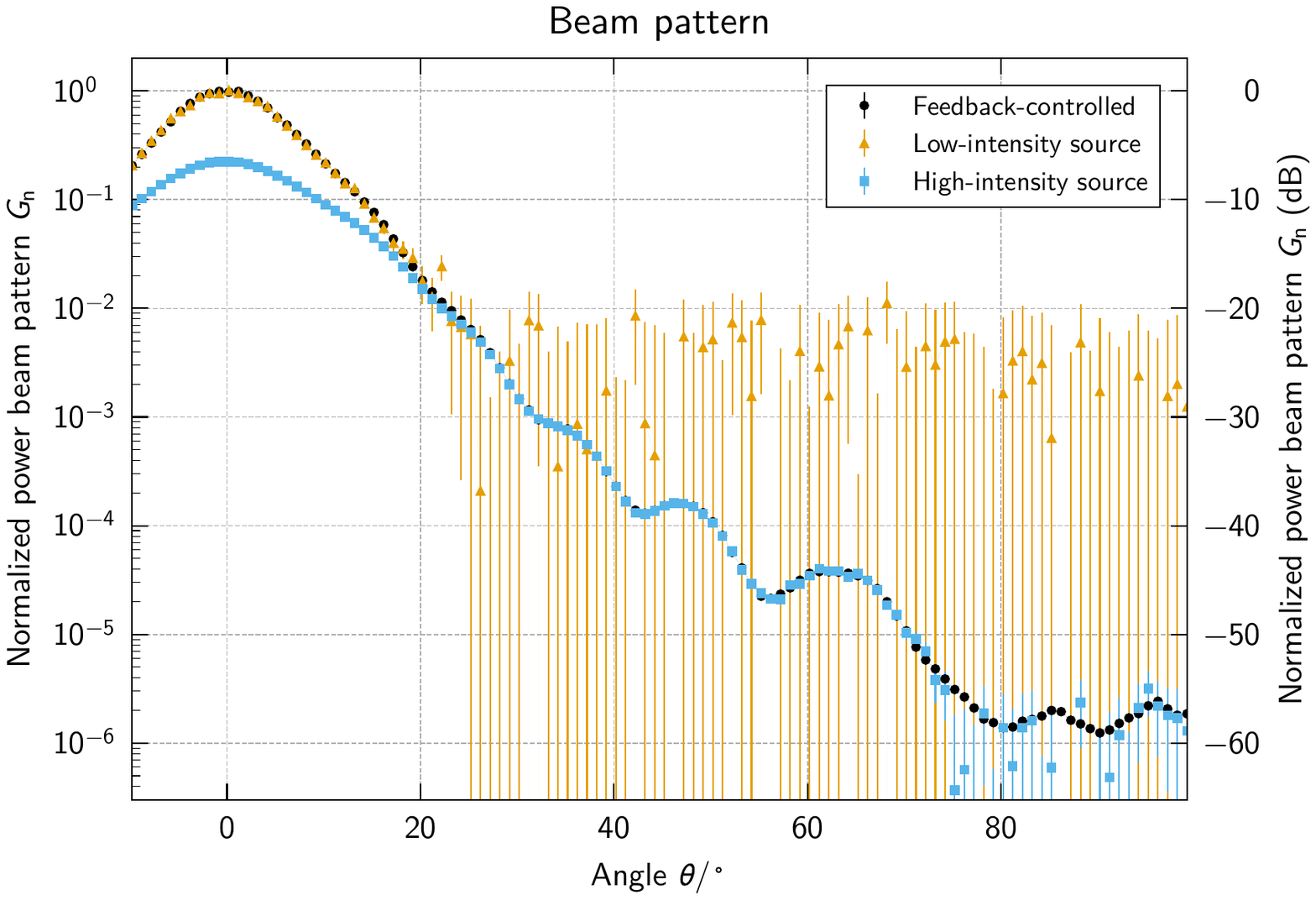}
    \caption{
        Comparison of the \textit{H}-plane beam patterns 
        of the W-band standard gain pyramidal horn antenna at \qty{81}{\GHz}
        measured using the feedback-controlled source (black circle), 
        the low-intensity static source (orange triangle), 
        and the high-intensity static source (sky blue square).
        The last one is not normalized at the beam center 
        but is scaled to the feedback-controlled measurement at $\theta = \ang{45}$
        because it is compressed at the main lobe due to the saturation of the detector under test.}
    \label{fig:beam_comparison}
\end{figure}

Figure \ref{fig:beam_comparison} shows the beam patterns 
measured with the low-intensity and high-intensity sources, 
compared to the result obtained using the feedback-controlled method.
The low-intensity source measurement achieved a dynamic range of only \qty{\sim 20}{\decibel}, 
which was insufficient to detect the side lobes.
The high-intensity source saturated the DUT around the main lobe 
(in the angular range of $|\theta| < \ang{30}$)
and caused a compression of up to \qty{6.5}{\decibel}.
Additionally, faint side lobes below \qty{-50}{\decibel} remained undetectable
because the low voltage gain of the DUT's preamplifier 
caused the noise floor to be dominated by their internal noise.
The feedback-controlled method effectively avoids
both low signal-to-noise ratio and distortion from saturation or nonlinearity.

\begin{figure}[!htbp]
    \centering
    \includegraphics[width=0.8\columnwidth]{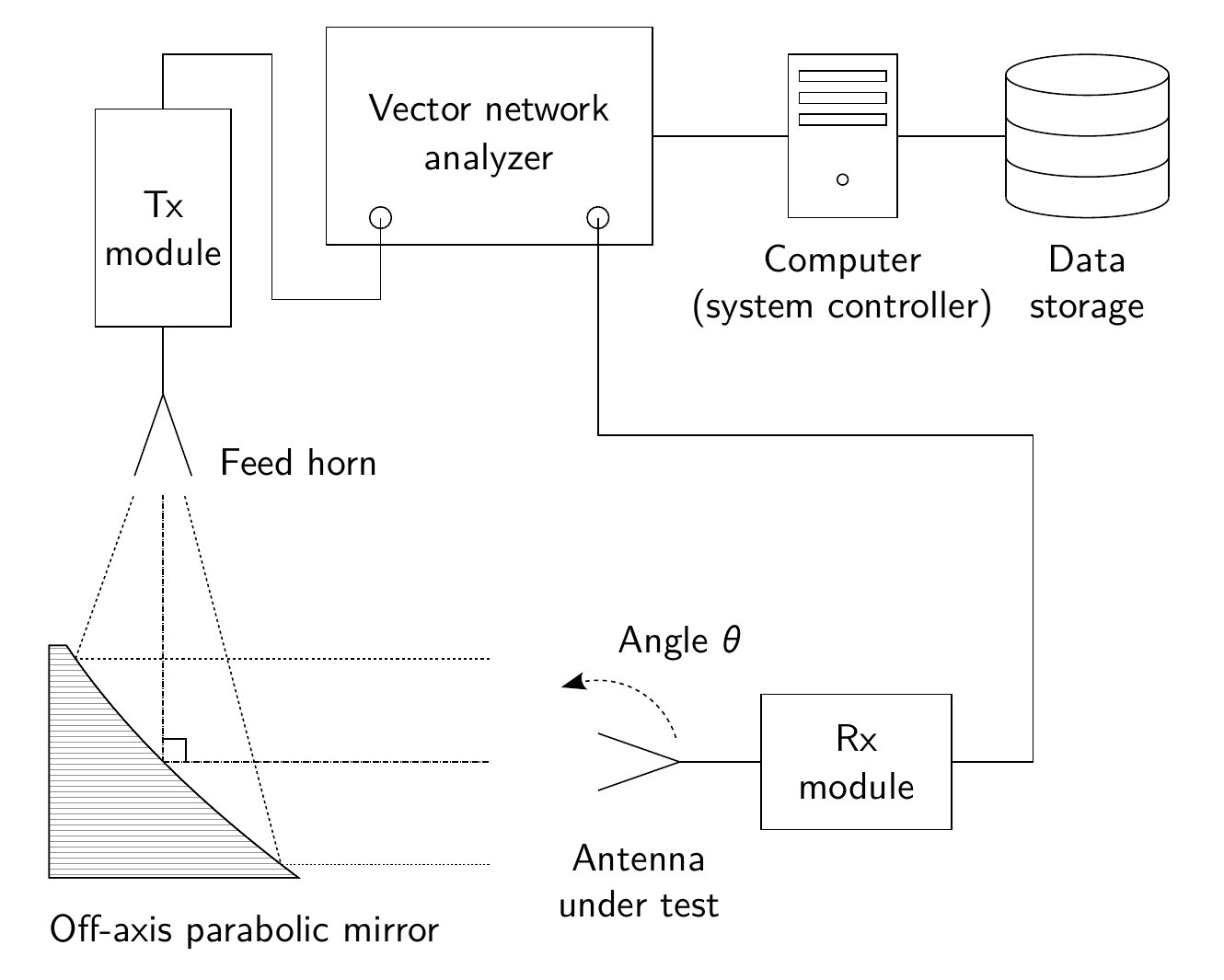}
    \caption{
        Block diagram of the beam pattern measurement system
        using the vector network analyzer (VNA).
        The frequency extender transmitter (Tx) module feeds the compact range with a microwave via the feed horn.
        The receiver (Rx) module receives the microwave through the antenna under test.
        The feed horn, optics geometry, and antenna under test are the same as the feedback-controlled system
        shown in Fig. \ref{fig:system}.
        The VNA measures $S_{21}$ (the scattering parameter from Tx to Rx)
        at every angular point.
    }
    \label{fig:system_vna} 
\end{figure}

We verified the feedback-controlled beam pattern measurement 
by evaluating its consistency with two reference measurements using conventional methods.
The first reference was a long-time integration measurement,
in which a low-intensity static source measurement with a longer signal integration time 
was repeated and averaged.
A signal integration time of \qty{100}{\second},
which is \num{10} times longer than in the previous measurement,
was allocated at each angular point,
and the measurement was repeated accordingly.
Due to time constraints, the measurements were sparsely done, at \ang{5.0} angular intervals.
A total of \num{1207} measurements were finally selected
based on the noise level and system temperature fluctuation. 
This corresponds statistically to a $\sqrt{\num{1207} \times \num{10}} = \qty{15.4}{\decibel}$
extension of the dynamic range compared to the previous measurement.
The second reference was a measurement using a vector network analyzer (VNA), 
a standard instrument in microwave engineering for characterizing microwave components.%
\footnote{
    VNAs are utilized for component- or optics-only beam pattern measurements,
    but are unavailable for measurements in TES-bolometer-installed telescopes.
    }
A separate test system was constructed, as shown in Fig. \ref{fig:system_vna}.
This system included a VNA (Keysight Technologies N5225B) 
along with a pair of W-band frequency extender transmitter and receiver modules 
(Virginia Diodes Inc. WR-10 VNAX).
The beam pattern was obtained by normalizing the angular pattern of $S_{21}$,
the scattering parameter from the transmitter to the receiver, computed by the VNA. 
The same feed horn, optical configuration, and AUT were used as in the feedback-controlled test system.

\begin{figure}[!htbp]
    \centering
    \includegraphics[width=0.90\columnwidth]{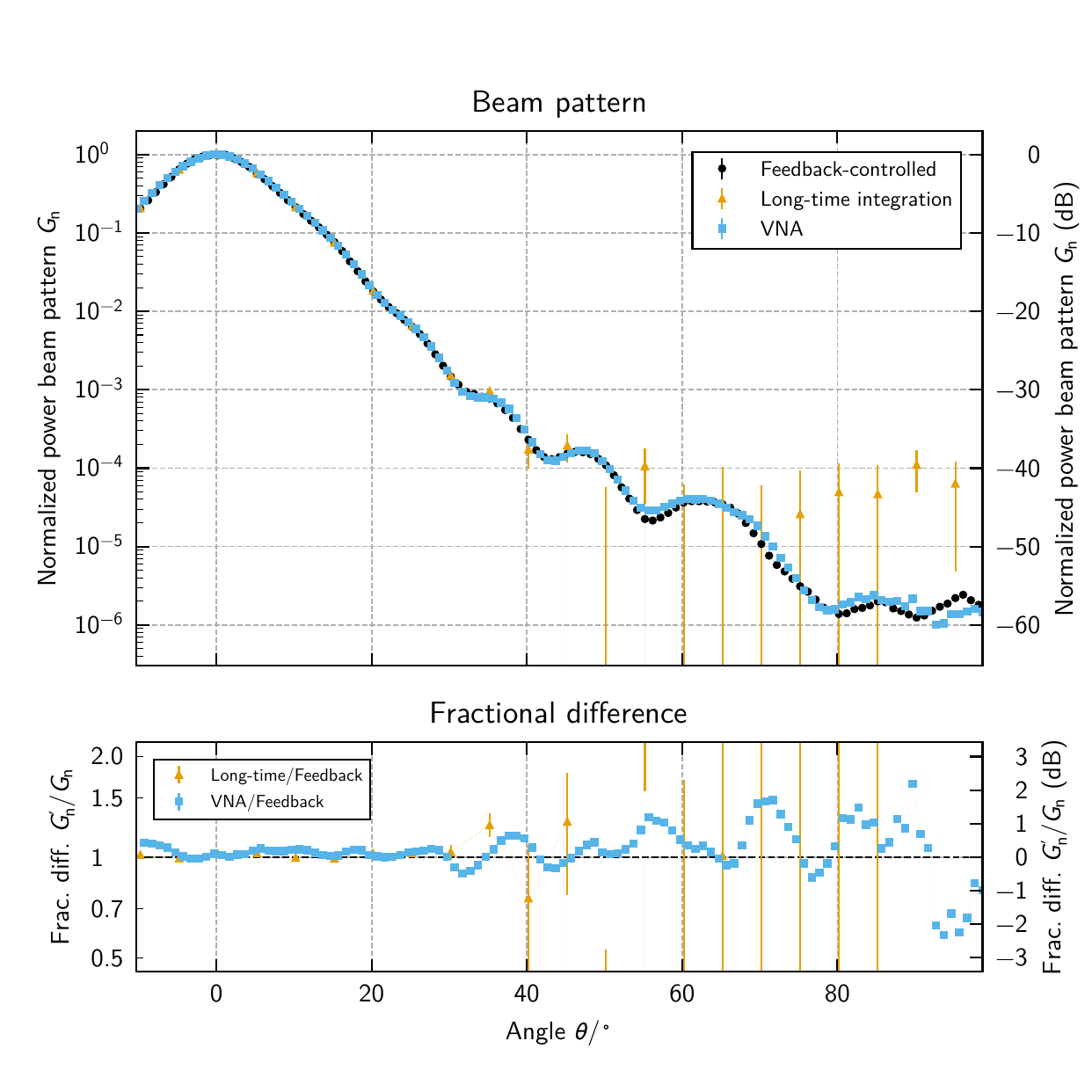}
    \caption{
        Comparison (top) and the fractional difference (bottom) 
        of the \textit{H}-plane beam patterns of the W-band standard gain pyramidal horn antenna at \qty{81}{\GHz}
        obtained from the feedback-controlled measurement (black circle), 
        the long-time integration measurement (orange triangle), 
        and the VNA measurement (sky blue square).
        The long-time integration measurement shows the unweighted, averaged beam pattern
        over \num{1207} repeats of the measurement with the low-intensity static source,
        each of which was conducted with a \num{10} times longer signal integration time
        and at \num{5} times sparser angular intervals than the previous measurement shown in Fig. \ref{fig:beam_comparison}.
        Its error bars show the standard error of the mean.
    }
    \label{fig:beam_consistency}
\end{figure}

Figure \ref{fig:beam_consistency} shows the comparison 
of the feedback-controlled measurement with the two reference measurements.
The fractional differences are also shown to evaluate their consistency.
The fractional difference between two beam patterns is defined as their fraction $G_\mathrm{n}^\prime / G_\mathrm{n}$,
which corresponds to their difference on a logarithmic scale.
The long-time integration measurement achieved a dynamic range of \qty{\sim 35}{\decibel} 
and agreed with the feedback-controlled measurement within $2\sigma$ uncertainty at most of the \ang{5}-step angular points.
However, the long-time integration measurement lacked sufficient angular resolution to resolve fine side lobe structures and a sufficient dynamic range to verify the faint side lobes.
In contrast, the VNA measurement provided the same angular resolution as the feedback-controlled measurement and a sufficiently high dynamic range to see faint side lobes.
It also showed good agreement with the feedback-controlled measurement 
at the \qty{\sim 2}{\decibel} level or better in terms of fractional difference.
The systematic difference at that level, particularly noticeable in the side lobe region,
is attributed to differences 
between the feedback-controlled measurement system and the VNA-based measurement system.
As the two measurements were conducted using different systems,
stray light, standing waves, and alignment are likely to differ, 
resulting in the slight disagreement.

\section{Conclusion}
\label{sec:conclusion}

In this study, we proposed a feedback-controlled beam pattern measurement method for CMB telescopes
that extends the dynamic range of the measured beam pattern
by using a power-variable calibration source.
As a proof of concept, 
we demonstrated the method by measuring the beam pattern of a horn antenna in a laboratory environment 
and showed that the measurement successfully obtained an additional dynamic range of \qty{60.3}{\decibel} 
attributed to the feedback control of the source power.
We also verified that the feedback-controlled measurement yielded results
consistent with those obtained using standard methods without feedback control.

The proposed method can be utilized to characterize the beam pattern of 
current- or next-generation CMB telescopes.
In particular, the method is expected to be powerful in laboratory testing of future CMB telescopes
since it can exclude the systematic effects caused by the TES bolometer's nonlinearity, 
which is of concern as discussed in Sec. \ref{sec:introduction},
from the beam pattern measurement.
Since TES bolometers have significantly lower saturation power 
than the diode amplitude detector used in this study, 
the transmitted power level may require adjustment.
This can be achieved by optimizing the source system,
for example, by inserting an attenuator at the feed horn.
Differences in optical efficiency
can also be canceled in a similar manner.
It should be noted that, unlike conventional methods,
the proposed method cannot simultaneously measure the beam patterns 
of multiple detectors within the same telescope,
as the feedback control operates on only one detector at a time.
In recent CMB telescopes, 
which have a large focal plane with \num{>1000} detectors,
we would perform feedback-controlled beam pattern measurements 
on some typical detectors, such as those at the center or edges of the focal plane,
rather than on all the detectors. 
These selective measurements are still valuable for characterizing telescope side lobes.
For instance, the measurements are useful 
to reveal the agreement of beam patterns between multiple pixels on the focal plane
and to validate the telescope's physical optical model.
In addition, the combination with measurements in conventional methods allows us 
to discuss potential systematic errors in the beam pattern characterization.
For example, we can verify whether the beam patterns measured using conventional methods are distorted
due to the nonlinearity of TES bolometers.

Beyond CMB telescope side lobe characterization, 
the method is applicable to a broad range of optical measurements
since the methodology we presented in Sec. \ref{sec:methodology} does not rely on any specific antenna design, detector type, or testing frequency.
As demonstrated in this study, 
it enables high-dynamic-range side lobe measurement of general antennas across a broad frequency range,
which was previously difficult to perform without a VNA. 
The same principle can be applied to other measurements, 
including band-pass, transmittance, and reflectance measurements 
of linear optical devices. 
By enabling the detection of faint nonidealities, 
the proposed high dynamic range approach enhances the understanding of the instrument's optical characteristics 
and contributes to the suppression of systematic errors in CMB B-mode polarization searches.

\section*{Acknowledgment}

This work was supported by
JSPS KAKENHI grant numbers 19KK0079, 20H01921, 22H04945, 25H00403, 
JSPS core-to-core program number JPJSCCA20200003,
and the World Premier International Research Center Initiative (WPI) of MEXT, Japan.
Calculations were performed on the KEK Central Computing System (KEKCC),
owned and operated by the Computing Research Center at KEK.
We thank Shogo Nakamura for giving useful suggestions on the experiments and data analysis.
We thank Shuhei Kikuchi for providing the temperature monitoring device for the test setup.
We thank Yuji Chinone for the useful discussion to improve the manuscript.
We thank Enago (www.enago.jp) for English language editing.

\vspace{0.2cm}
\noindent

\let\doi\relax


\begin{thebibliography}{10}

\bibitem{Seljak1997}
    U.~Seljak and M.~Zaldarriaga, Phys. Rev. Lett. {\bf 78}, 2054--2057 (1997) [arXiv:astro-ph/9609169].\\
    \doi{https://doi.org/10.1103/PhysRevLett.78.2054}

\bibitem{Kamionkowski1997}
    M.~Kamionkowski, A.~Kosowsky, and A.~Stebbins, Phys. Rev. Lett. {\bf 78}, 2058--2061 (1997) [arXiv:astro-ph/9609132].
    \doi{https://doi.org/10.1103/PhysRevLett.78.2058}

\bibitem{SPT3G2018}
    A.~J.~Anderson et al., J. Low Temp. Phys. {\bf 193}(5–6), 1057–1065 (2018).
    \doi{https://doi.org/10.1007/s10909-018-2007-z}

\bibitem{SO2019}
    P.~Ade et al. [The Simons Observatory collaboration], J. Cosmol. Astropart. Phys. {\bf 2019}(02), 056 (2019) [arXiv:1808.07445 [astro-ph]].
    \doi{https://doi.org/10.1088/1475-7516/2019/02/056}

\bibitem{BK2022}
    P.~A.~R. Ade et al. [BICEP/\textit{Keck} Collaboration], Astrophys. J. {\bf 927}(1), 77 (2022) [arXiv:2110.00482 [astro-ph]].
    \doi{https://doi.org/10.3847/1538-4357/ac4886}

\bibitem{CMBS42022}
    K.~Abazajian et al. [The CMB-S4 Collaboration], Astrophys. J. {\bf 926}(1), 54 (2022) [arXiv:2008.12619 [astro-ph]].
    \doi{https://doi.org/10.3847/1538-4357/ac1596}

\bibitem{LiteBIRD2023}
    E.~Allys et al. [LiteBIRD Collaboration], Prog. Theor. Exp. Phys. {\bf 2023}(4), 042F01 (2022) [arXiv:2202.02773 [astro-ph]].
    \doi{https://doi.org/10.1093/ptep/ptac150}

\bibitem{Stevens2020}
    J.~R.~Stevens et al., J. Low Temp. Phys. {\bf 199}(3–4), 672–680 (2020) [arXiv:1912.00860 [astro-ph]].\\
    \doi{https://doi.org/10.1007/s10909-020-02375-9}

\bibitem{Jaehnig2020}
    G.~C.~Jaehnig et al., J. Low Temp. Phys. {\bf 199}(3–4), 646–653 (2020).
    \doi{https://doi.org/10.1007/s10909-020-02425-2}

\bibitem{deHaan2024}
    T.~de~Haan, Proc. SPIE 13102, Millimeter, Submillimeter, and Far-Infrared Detectors and Instrumentation for Astronomy XII, 1310208 (2024) [arXiv:2406.19567 [astro-ph]].
    \doi{https://doi.org/10.1117/12.3018503}

\bibitem{Dutcher2024}
    D.~Dutcher et al., J. Low Temp. Phys. {\bf 214}(3–4), 247–255 (2024) [arXiv:2311.05583 [astro-ph]].\\
    \doi{https://doi.org/10.1007/s10909-023-03045-2}

\bibitem{WMAP2003a}
    L.~Page et al., Astrophys. J. Suppl. {\bf 148}(1), 39–50 (2003) [arXiv:astro-ph/0302214].\\
    \doi{https://doi.org/10.1086/377223}

\bibitem{PB2014}
    P.~A.~R.~Ade et al. [The Polarbear Collaboration], Astrophys. J. {\bf 794}(2), 171 (2014) [arXiv:1403.2369 [astro-ph]].
    \doi{https://doi.org/10.1088/0004-637x/794/2/171}

\bibitem{Planck2016}
    R.~Adam et al. [Planck Collaboration], Astron. Astrophys. {\bf 594}, A7 (2016) [arXiv:1502.01586 [astro-ph]].
    \doi{https://doi.org/10.1051/0004-6361/201525844}

\bibitem{WMAP2003b}
    C.~Barnes et al., Astrophys. J. Suppl. {\bf 148}(1), 51–62 (2003) [arXiv:astro-ph/0302215].\\
    \doi{https://doi.org/10.1086/377227}

\bibitem{QUIET2013}
    C.~Bischoff et al. [QUIET Collaboration], The Astrophys. J. {\bf 768}(1), 9 (2013) [arXiv:1207.5562 [astro-ph]].
    \doi{https://doi.org/10.1088/0004-637X/768/1/9}

\bibitem{CLASS2024}
    R.~Datta et al., Astrophys. J. Suppl. {\bf 273}(2), 26 (2024) [arXiv:2308.13309 [astro-ph]].\\
    \doi{https://doi.org/10.3847/1538-4365/ad50a0}

\bibitem{BK2015}
    P.~A.~R.~Ade et al. [The BICEP2 and Keck Array Collaborations], Astrophys. J. {\bf 806}(2), 206 (2015) [arXiv:1502.00596 [astro-ph]].
    \doi{https://doi.org/10.1088/0004-637X/806/2/206}

\bibitem{Stutzman1998}
    W.~L.~Stutzman and G.~A.~Thiele, \emph{Antenna Theory and Design} (Wiley, New York, 1998), 2nd ed.

\end{thebibliography}
\end{document}